\documentstyle{mn}

\input epsf
\begin{document}
\journal{astro-ph/0105361~~~~~LAPTH~847/01} 
\title[The lepton asymmetry and cosmology]{The lepton asymmetry: the
  last chance for a critical-density cosmology?}
\author[J.~Lesgourgues and A.~R.~Liddle]{Julien Lesgourgues$^1$ and
  Andrew R.~Liddle$^2$\\
$^1$LAPTH, Chemin de Bellevue, B.P. 110, 74941 Annecy-le-Vieux Cedex, 
France.\\
$^2$Astronomy Centre, University of Sussex, Falmer, Brighton BN1
  9QJ, United Kingdom.}  
\maketitle
\begin{abstract}
We use a wide range of observations to constrain cosmological models
possessing a significant asymmetry in the lepton sector, which offer
perhaps the best chance of reconciling a critical-density Universe
with current observations. The simplest case, with massless neutrinos,
fails to fit many experimental data and does not lead to an acceptable
model. If the neutrinos have mass of order one electron-volt (which is
favoured by some neutrino observations), then models can be
implemented which prove a good fit to microwave anisotropies and
large-scale structure data. However, taking into account the latest
microwave anisotropy results, especially those from Boomerang, we show
that the model can no longer accommodate the observed baryon fraction
in clusters. Together with the observed acceleration of the present
Universe, this puts considerable pressure on such critical-density
models.
\end{abstract}

\section{Introduction}

The recent use of the magnitude--redshift relation of type Ia
supernovae to infer that the present Universe is accelerating
(Perlmutter et al.~1998, 1999; Schmidt et al.~1998; Riess et al.~1999)
has led to a consensus that the cosmological model best fitting
current data is a spatially-flat cold dark matter Universe with a
matter density around 0.3 of the critical-density (Peebles 1984;
Turner, Steigman \& Krauss 1984; Efstathiou, Sutherland \& Maddox
1990). This model, known as $\Lambda$CDM, can boast an impressive
range of observational successes, with its main drawback being
theoretical objection both to the magnitude and the required recent
dominance of the cosmological constant term.

It is often stated that while the supernova results are powerful in
themselves, it is unlikely that they would have been widely accepted
had there not been considerable other evidence pointing towards this
favoured cosmology (Krauss \& Turner 1995; Ostriker \& Steinhardt
1995). Amongst that, one might mention the shape of the galaxy
correlation function, the combination of the cluster baryon fraction
with standard Big Bang Nucleosynthesis (BBN), and the flat geometry
inferred from the cosmic microwave background combined with the low
matter density implied by direct observations.

In this paper, we examine the extent to which these additional
arguments might be undermined in an alternative cosmological model,
which features an asymmetry in the lepton sector leading to a higher
than usual abundance of neutrinos. It was claimed recently 
(Adams \& Sarkar 1998, Lesgourgues \& Peloso 2000) that these
models offer one of the best remaining prospects for salvaging the
idea of a critical-density Universe [the other main option being the
Broken Scale Invariance models (Barriga et al.~2000)], and although
the likelihood of doing so is small it is judicious to be aware of the
possibility in order to balance its drawbacks with those of the
cosmological constant model.

The lepton asymmetry model relies on primordial processes to create an
imbalance between the numbers of neutrinos and antineutrinos in the
Universe, which may reside in any of the three neutrino families. This
would be the leptonic analogue of the (presently unknown) processes
leading to the baryon number of the Universe, though in this case
interesting effects only arise for an asymmetry of order one, whereas
the baryon-to-photon ratio is of order $10^{-9}$. There are many
particle physics motivated scenarios for generating such a large
lepton asymmetry (e.g. Foot, Thomson \& Volkas~1996; Casas, Cheng \&
Gelmini~1999; March-Russell, Murayama \& Riotto~1999; McDonald~2000;
Dolgov et al.~2000; Kirilova \& Chizhov~2000; Di Bari \& Foot~2001).
The lepton asymmetry leads to two important physical effects. The
first is that it modifies standard nucleosynthesis calculations, since
the neutrino asymmetry alters the initial balance of protons and
neutrons, and it has been known for some time that matching the
element abundances in the presence of a strong lepton asymmetry can
require a higher baryon fraction than standard nucleosynthesis (see
for instance Kang \& Steigman~1992; Esposito et al.~2000, 2001;
Kneller et al.~2001).  The second is that it increases the radiation
density in the Universe, by boosting the neutrino density beyond its
usual value of 0.68 times the photon density.

At face value, these have highly desirable implications in cosmology
for those favouring critical density on grounds of elegance, as was
first pointed out by Adams \& Sarkar (1998). First, if the preferred
baryon density from nucleosynthesis could be significantly increased
through the leptonic asymmetry, the cluster baryon fraction would then
become a strong argument for critical density rather than
against. Further, the extra radiation leads to a delay in
matter--radiation equality, which shifts the characteristic bend in
the matter power spectrum to larger scales mimicking the effect of the
reduced matter density in the $\Lambda$CDM model. Finally, it was
stressed (Lesgourgues \& Peloso~2000; Esposito et al.~2001; Kneller et
al.~2001) that a leptonic asymmetry could help in explaining last
year's microwave anisotropy results from Boomerang (de Bernardis et
al.~2000) and Maxima (Hanany et al.~2000), in which the observed
weakness of the second acoustic peak favoured a high baryon
fraction. Indeed, our initial studies for this present work indicated
that critical-density models with leptonic asymmetry could fit not
only these data, but also up-to-date constraints from
large-scale-structure, the cluster baryon fraction, and primordial
element abundances. So, although unable to explain the supernovae
data, this model could undermine much of the other evidence supporting
the $\Lambda$CDM model.

This picture seems to be less promising after the publication of new
microwave anisotropy results by DASI (Halverson et al.~2001; Pryke et
al.~2001) and updated data analysis by Boomerang (Netterfield et
al.~2001) and Maxima (Lee et al.~2001), which are incorporated into
the results reported here. The new results contain no direct
independent evidence for a cosmological constant, but within the
framework of $\Lambda$CDM models exhibit excellent agreement with the
baryon fraction from standard nucleosynthesis, while an excess of
baryons is a key ingredient for the success of critical-density
models. However, a high baryon fraction may yet be allowed in the
presence of a leptonic asymmetry, and the true test of the idea lies
in detailed comparison with observations, which is our purpose in this
paper.  We will see that the situation is not at all promising for the
simplest case of massless neutrinos, which have trouble fitting many
types of observation. However, there is now considerable experimental
evidence that neutrinos actually possess a small mass, and the effects
of this need to be included. The neutrino mass provides an additional
modification to the matter power spectrum through neutrino
free-streaming (as in the mixed dark matter scenario), and we find
that this enables excellent fits to many observational data to be
obtained. Unfortunately, due to the latest Boomerang results, the
model fails to explain the baryon fraction in clusters, as well as the
present acceleration.

\section{The lepton asymmetry models}

The lepton asymmetry models are in most respects the same as
conventional structure formation models, in particular relying on the
presence of cold dark matter, but add new parameters describing the
magnitude of the lepton asymmetry and the mass of the neutrinos. In
principle the masses at least are not extra parameters as compared to
the standard cosmology, in that there is now substantial evidence that
neutrinos do have mass; however in the presence of a lepton asymmetry
the neutrinos may have a more significant impact on predictions for a
given mass as the asymmetry increases their number density. In
compensation for adding these extra parameters, we remove the
cosmological constant.

Provided that neutrinos reached thermal equilibrium before decoupling,
the leptonic asymmetry for each flavour species can be conveniently
parametrized by the ratio of chemical potential over temperature,
$\xi_{\nu_i} = \mu_{\nu_i} / T_{\nu_i}$ (with $i \in \{e, \mu,
\tau\}$).  Neutrinos with a chemical potential are called degenerate
neutrinos, because the asymmetry enhances their total density. When
the neutrinos are in the relativistic regime, this effect is strictly
equivalent to a change in the effective number of standard neutrinos,
in excess of the usual value of 3, of
\begin{equation} 
\label{degeneracy}
\Delta N_{\rm eff}=\sum_i \left[ 30 (\xi_{\nu_i} / \pi)^2/7 +
15(\xi_{\nu_i} /\pi)^4/7 \right]~.
\end{equation}

All the generation mechanisms proposed so far predict different
$\xi_{\nu_i}$'s for each species, at least in absence of
fine-tuning. This is a crucial point because the density of $\nu_e$
and $\nu_{\mu} + \nu_{\tau}$ have opposite effects on the
neutron-to-proton ratio at freeze-out during BBN, and on the
production of light elements. More precisely, nucleosynthesis in the
presence of a lepton asymmetry (known as degenerate Big Bang
Nucleosynthesis) requires three ingredients in order to be compatible
with the observed abundances of deuterium, helium-4 and lithium-7 :
(i) an increase\footnote{Actually, there is also a small allowed
region in parameter space with $\xi_{\nu_e} < 0$, reduced baryon
density and $N_{\rm eff} < 3$, but this is irrelevant in the present
framework.}  in $\nu_e$ density ($\xi_{\nu_e} > 0$); (ii) an increase
in the baryon density; (iii) an increase in the total density of
radiation (and expansion rate of the Universe), bigger than the one
resulting from (i), and parametrized by an effective number of
standard neutrino species $N_{\rm eff} > 3$. There are many
possibilities for enhancing the radiation density: the mu and/or tau
neutrino may have a leptonic asymmetry bigger than that of the
electronic neutrino,\footnote{i.e., $|\xi_{\nu_{\mu}} +
\xi_{\nu_{\tau}}| > \xi_{\nu_e}$. However, successful implementations
require only a factor of order five or so between $|\xi_{\nu_{\mu}} +
\xi_{\nu_{\tau}}|$ and $\xi_{\nu_e}$, which seems to be compatible
with most mechanisms of large leptonic asymmetry generation.}  or may
become slightly non-relativistic during BBN (Hansen \& Villante 2000),
and apart from the three flavour neutrinos, many scenarios predict
that extra relativistic degrees of freedom could be present during
nucleosynthesis (for instance, axions). Hannestad (2001) has recently
studied limits on the number of neutrino species from the latest data
(including models with a cosmological constant).

Standard BBN, which corresponds to $\xi_{\nu_i}=0$, predicts a baryon
fraction given by $\Omega_{\rm b} h^2 = 0.019 \pm 0.002$, and an
effective neutrino number close to 3. In the following analysis we
will focus on a baryon fraction in the range $0.015 <\Omega_{\rm b}
h^2 < 0.035$. In this case, according to the most recent studies of
degenerate BBN (Esposito et al.~2000, 2001; Kneller et al.~2001), the
$\nu_e$ asymmetry parameter should be in the range $0 < \xi_{\nu_e} <
0.5$, while the required effective neutrino number could be as big as
15 or even 20. Using equation (\ref{degeneracy}), we see that the
contribution of $\xi_{\nu_e}$ to $\Delta N_{\rm eff}$ is negligible;
therefore, when studying the spectrum of microwave anisotropies and
large-scale structure, we can forget completely about $\xi_{\nu_e}$,
and consider the constraint from degenerate BBN to lie simply in
the $(\Omega_{\rm b} h^2, N_{\rm eff})$ plane.

The calculations of matter and radiation power spectra were carried out
using a modified version of the CMBFAST code (Seljak \& Zaldarriaga
1996), as described in Lesgourgues \& Pastor (1999). 

\section{Observational constraints}

For each model, we define a $\chi^2$ statistic including the following
terms: first, 19 data points from the new analysis of Boomerang
(Netterfield et al.~2001), 13 from the new analysis of Maxima (Lee et
al.~2001), and 9 from DASI (Halverson et al.~2001); second, 22 data
points from the PSCz redshift survey (Hamilton \& Tegmark 2000); and
finally, a constraint on the matter spectrum normalization $\sigma_8$
from the number density of galaxy clusters. For the last, we adopt the
rather conservative constraint $\sigma_8 = 0.56 \pm 0.056$
(1-$\sigma$) from Viana \& Liddle (1999). We also reran the analysis
using the tighter limit $\sigma_8 = 0.495 \pm 0.034$ (1-$\sigma$)
recently obtained by Pierpaoli, Scott \& White (2000), but this made
no qualitative difference to our conclusions and so we do not report
those results here.

For Boomerang and Maxima, we treat each data point as uncorrelated,
with approximately gaussian window functions.  We take into account
the fully correlated calibration error and the multipole-dependent
beam plus pointing error, minimizing over the corresponding
parameters. The likelihood function can be written as ${\cal L}=e^{-
\chi^2 / 2}$ provided that the data points are almost gaussian
distributed. For Maxima we use the results for $\Delta T$ (Table 1 in
Lee et al.~2001) which have almost symmetric errors (except for the
last two points, which doesn't matter because these points provide
mainly upper limits). The beam plus pointing uncertainty (calculated
from the same Table 1) turns out also to be symmetric for $\Delta
T_l$, so we can define a $\chi^2$ for Maxima as
\begin{equation}
\chi^2 = \sum_l \frac{(\Delta T_l^{\rm theo} - 
(1 + c~\sigma_c + b~\sigma_{b,l}) \Delta T_l^{\rm obs})^2}{\sigma_l^2}
+ b^2 + c^2 \,, \;\,
\end{equation}
with a 1-$\sigma$ calibration uncertainty $\sigma_c=0.04$, and a
1-$\sigma$ beam plus pointing uncertainty which is well fitted by the
function $\sigma_{b,l} = 10^{-6} l^{1.7}$.  For each model, we
minimize the $\chi^2$ over $b$ and $c$.  For Boomerang, we use a
similar expression. However, in Netterfield et al.~(2001), symmetric
error bars are given for $D_l \equiv (\Delta T_l)^2 = l(l+1) C_l / 2
\pi$.  Therefore, we define the $\chi^2$ directly on this quantity,
with a 1-$\sigma$ calibration uncertainty $\sigma_c=0.20$. The beam
errors (read from figure~2 in Netterfield et al.~2001) are symmetric for
$\Delta T_l$, with $\sigma_{b,l} = 0.215 \times 10^{-6} l^2$ at
1-$\sigma$.  For simplicity, we assume a gaussian beam error
for $D_l$, with twice the uncertainty.

For the DASI data, Pryke et al.~(2001) indicate that the use of the
exact window functions, and of a transformation that gives exactly
gaussian errors (Bond, Jaffe \& Knox 2000), has only a modest impact
on parameter extraction. On the other hand, the points cannot be
treated as uncorrelated. Accordingly, we define the following
covariance matrix
\begin{equation}
M_{ij}= \Delta D_i ~V_{ij} ~\Delta D_j + s^2 D_i D_j \,,
\end{equation}
using the data points $D_l \pm \Delta D_l$ and the correlation matrix
$V_{ij}$ from Halverson et al.~(2001, Tables I and II).  The
fully-correlated uncertainty $s$ equals 0.08. The $\chi^2$ is then
defined as
\begin{equation}
\chi^2 = \sum_{i,j} (D_i^{\rm theo} - D_i^{\rm obs}) M_{ij}^{-1}
(D_j^{\rm theo} - D_j^{\rm obs}) \,.
\end{equation}

We compute the total $\chi^2$ values on a grid in parameter space, and
perform a multidimensional cubic spline interpolation in order to find
the minimum and the confidence limits on each
parameter.\footnote{To marginalize over unwanted parameters, we
maximize the likelihood function instead of integrating over these
parameters (the two techniques would be strictly equivalent only for a
multivariate gaussian likelihood). Therefore, our confidence limit
computation scheme is less rigorous than in current state-of-the-art
analyses (Lange et al.~2000; Jaffe et al.~2001; Netterfield et
al.~2001), but because of its simplicity it is widely used by many
other authors, and gives a fairly good hint of the true error bars [see
for instance the discussion in Tegmark \& Zaldarriaga (2000) and
Tegmark, Zaldarriaga \& Hamilton (2001)].} The precision and
efficiency of this method depends crucially on the choice of a
particular parameter basis.  Our seven cosmological parameters are the
overall normalization (adjusted automatically to match the COBE
observations (Bennett et al.~1996; Bunn \& White 1997) by the CMBFAST
code that we use), the baryon fraction $\Omega_{\rm b} h^2$, the
scalar tilt $n_s$, the optical depth to reionization $\tau$, the mass
of the degenerate neutrino $m_{\nu}$, the effective neutrino number
$N_{\rm eff}$ (we recall that in the case of massive neutrinos, this
number is defined at nucleosynthesis, not today), and a final
parameter measuring the dark matter density. This last parameter could
be taken as $\Omega_{\rm dm} h^2$ (where $\Omega_{\rm dm} =
\Omega_{\rm cdm} + \Omega_{\nu}$); however, this choice would lead to
an exceedingly large computing time because there is a degeneracy
between $\Omega_{\rm dm} h^2$ and the neutrino parameters ($m_{\nu}$,
$N_{\rm eff}$). In other words, for a given cosmological model and set
of observations, the likelihood regions are elongated along a
direction that can be found only empirically, and the $\chi^2$ varies
very slowly when the function $p_h(\Omega_{\rm dm} h^2, N_{\rm eff},
m_{\nu})$ associated with the degeneracy is almost constant.  The best
time-saving strategy is to directly use $p_h$ as the last free
cosmological parameter. The preferred value of $h$ (and of any other
combination of the cosmological parameters) can then be recovered a
posteriori.  In most cases studied hereafter, we find that
$p_h=\Omega_{\rm dm} h^2 N_{\rm eff}^{-1} (3.5 + m_{\nu})^{-1}$ is a
fairly good parametrization of the degeneracy in the vicinity of the
minimum (with $m_{\nu}$ expressed in electron-volts).

In addition to the six free cosmological parameters, our model
includes a free PSCz bias. The number of degrees of freedom is therefore 
$(19 + 13 + 9 + 22 + 1) - (6 + 1) = 57$.  Actually, the constraints from 
PSCz on the largest scales are so loose that this number is somewhat
overestimated.

\begin{figure}
\centering 
\leavevmode
\epsfxsize=2.7cm
\epsfbox{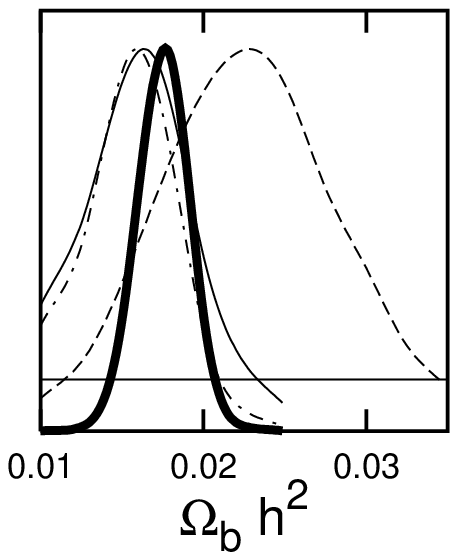}
\epsfxsize=2.7cm
\epsfbox{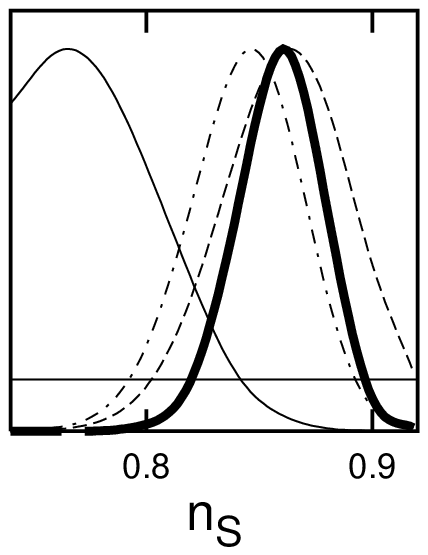}
\epsfxsize=2.7cm
\epsfbox{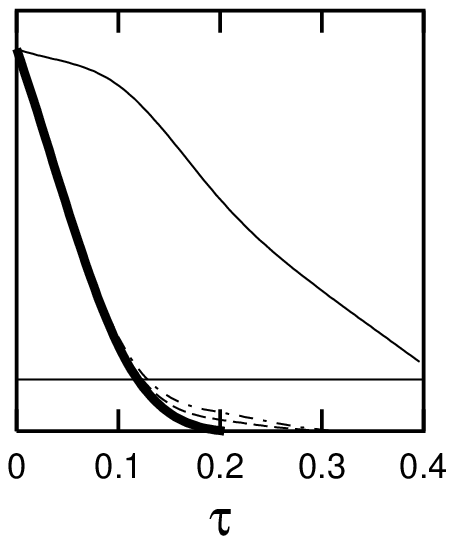}
\vspace{0.5cm}
\\
\leavevmode
\epsfxsize=2.7cm
\epsfbox{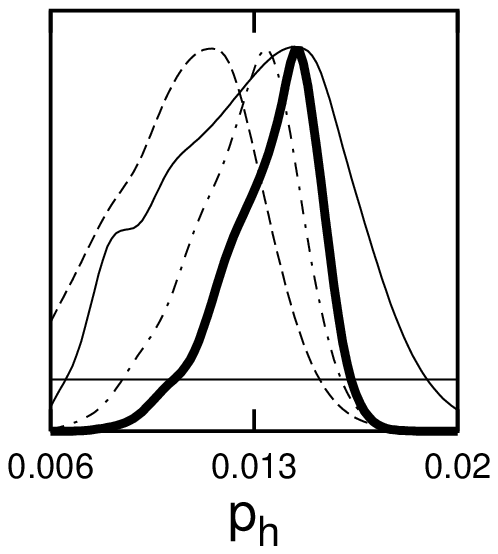}
\epsfxsize=2.7cm
\epsfbox{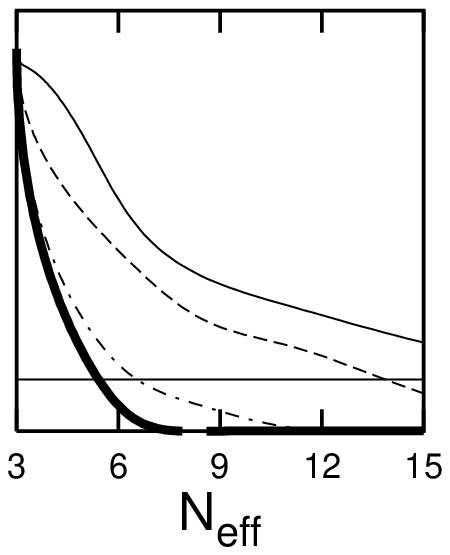}
\hspace*{2.5cm} \mbox{~}
\\
\caption[fig1]{\label{fig1} The probability distribution for each
cosmological parameter, in the case of one massless degenerate
neutrino. The thick solid lines show the result including all the
data. Parameter values are allowed at the 95 per cent confidence level
when the probability exceeds the horizontal line.  The thin curves
show the constraints obtained by combining just one CMB experiment
with the other non-CMB data: solid is Boomerang, dashed is Maxima and
dot-dashed is DASI. Although Maxima favours a significantly higher
baryon fraction, and Boomerang a lower scalar tilt, the three data
sets are found to be perfectly compatible.}
\end{figure}

\subsection{The massless neutrino case}

The results for massless neutrinos are summarized by the individual
parameter probability distributions, plotted in figure~\ref{fig1}.
The values for the best-fitting model and the 95 per cent confidence
level are given in Table~\ref{tab1}.

The results for this model are quite disappointing. The best-fitting
model has a $\chi^2$ of 61, which given the number of degrees of
freedom looks quite satisfactory. However the properties of the
best-fitting model are quite undesirable.  The preferred values
$\Omega_{\rm b} h^2=0.018$ and $N_{\rm eff}=3$ match the standard BBN
prediction, but this is not a good thing in the present context; since
we don't obtain a high baryon density, we cannot explain the cluster
baryon fraction (our predicted value for $\Omega_{\rm b} h^{1.5}$ and
the observed lower bound (Ettori \& Fabian 1999) have no overlap at
the 2-$\sigma$ level).  A further problematic aspect of this model is
its low preferred value of $h$: this gives an impressively large age,
but is in considerable discrepancy with direct $h$
measurements. Finally, we must recall that we have made no attempt to
obtain a presently accelerating universe.

We are therefore forced to conclude that the massless neutrino case
has too many failings against observations to be considered a viable
model.

\begin{table}
\begin{tabular}{|l|lll|lll|}
\hline 
& \multicolumn{3}{c}{Massless $\nu$}
& \multicolumn{3}{c}{Massive $\nu$}
\\
 & min & best & max & min & best & max
\\
\hline
$\Omega_{\rm b} h^2$ &0.014&0.018&0.021&0.016&0.020&0.025 \\
$n_s$ &0.82&0.86&0.90&0.90&0.96&1.02 \\
$\tau$ &0&0&0.12&0&0.10&0.36 \\
$p_h$ &0.010&0.014&0.016&0.0066&0.0075&0.0089 \\
$N_{\rm eff}$ &3&3&5.5&7&11&15 \\
$m_{\nu} ({\rm eV})$ &--&--&--&0.60&0.85&1.5 \\
$b$ &1.05&1.2&1.35&1.2&1.4&1.5 \\
\hline
$h$ &0.38&0.41&0.47&0.54&0.60&0.71 \\
$\Omega_{\rm cdm} h^2$ &0.13&0.15&0.17&0.22&0.29&0.42 \\
$\Omega_{\nu} h^2$ &--&--&--&0.03&0.06&0.08 \\
$\xi_{\nu} $ &0&0&2.1&2.8&3.3&3.9 \\
$\sigma_8 $ &0.57&0.65&0.73&0.45&0.54&0.63 \\
$t_0 ({\rm Gyr})$ &14&16&17&9&11&13 \\
$\Omega_{\rm b} h^{1.5}$ &0.022&0.028&0.034&0.023&0.026&0.033 \\
\hline
\end{tabular}
\caption[tab1]{\label{tab1} The preferred value and 95 per cent
confidence limits for each cosmological parameter, in the case of one
massless and one massive degenerate neutrino family. The upper part
refers to the parameter basis used in the interpolation, with $p_h$
defined as in the text. The lower part refers to useful combinations
of these parameters, including the leptonic asymmetry parameter
$\xi_{\nu}$, the age of the Universe $t_0$ and the quantity
$\Omega_{\rm b} h^{1.5}$ which can be compared with the baryon
fraction in galaxy clusters (the range given by Ettori \& Fabian
(1999) is $0.060 \pm 0.025$ at 2-$\sigma$ confidence level).}
\end{table}

\subsection{The massive neutrino case}

We now suppose that the neutrino family with the leptonic
asymmetry\footnote{In the massless case our results were
model-independent, since we did not privilege a particular scenario
for the origin of the extra relativistic degrees of freedom. When
taking into account a neutrino mass $m_{\nu}$, we could distinguish
various cases: first, the large chemical potential responsible for
$N_{\rm eff} > 3$ during BBN can belong to the massive neutrino
family; alternatively, it can be shared between one species with
negligible mass and one with arbitrary mass $m_{\nu}$; finally, the
extra radiation density could be attributed to particles other than
flavour neutrinos. For brevity, we only discuss the simplest case of
a single massive degenerate neutrino family ($\nu_{\mu}$ or
$\nu_{\tau}$). Most other situations would give comparable results,
but with a higher preferred value of the mass, since the neutrino
free-streaming effect is enhanced by the leptonic asymmetry
(Lesgourgues \& Pastor 1999).}  has a mass $m_{\nu}$. Now the
degenerate neutrino can make up a significant fraction of the dark
matter, as in the mixed dark matter scenario, and its free-streaming
while relativistic suppresses short-scale matter perturbations.  With
this additional free parameter, the minimum of $\chi^2$ shows a large
degeneracy along $p_h$: unreasonably large values of $h$ are allowed,
with a huge effective neutrino number maintaining the first acoustic
peak and the power spectrum with the right shape and amplitude.  This
parameter region is uninteresting and should be removed.  Indeed,
$h>0.58$ corresponds to a Universe younger than $t_0 = 11$~Gyr, which
is almost completely excluded. So, we must add to the $\chi^2$ a
``weak age prior'' $t_0 \geq 11$~Gyr.\footnote{Technically, this is
done by multiplying the likelihood function by a gaussian cut-off, if
and only if $t_0 \leq 11$~Gyr. The variance is chosen so that at $t_0
= 10$~Gyr the cut-off factor equals 1/2.} It is important to note that
this prior almost does not affect the goodness-of-fit of the model, since 
the best-fitting model has $t_0$ close to 11~Gyr anyway.

\begin{figure}
\centering 
\leavevmode 
\epsfxsize=2.7cm 
\epsfbox{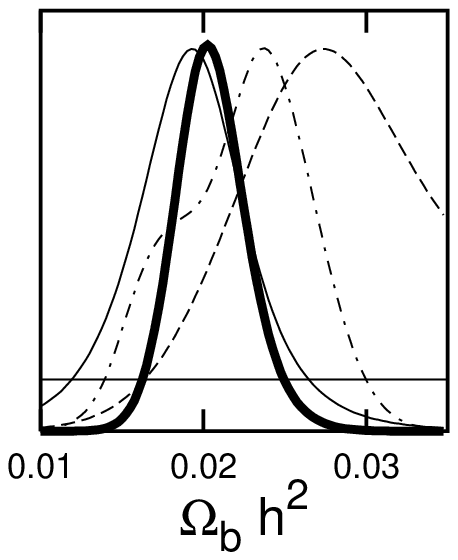}
\epsfxsize=2.7cm 
\epsfbox{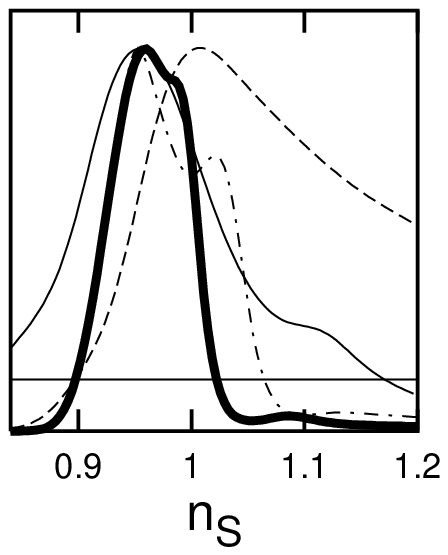}
\epsfxsize=2.7cm 
\epsfbox{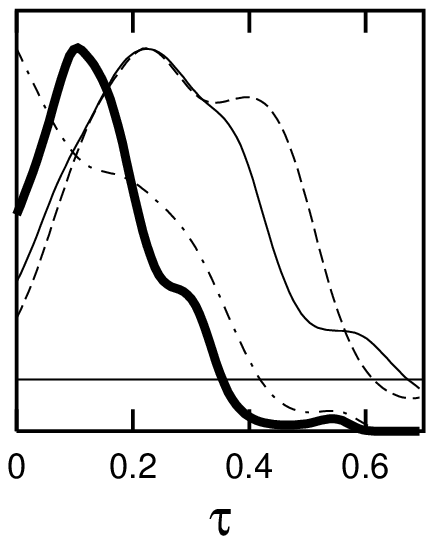}
\vspace{0.5cm}
\\
\leavevmode
\epsfxsize=2.7cm
\epsfbox{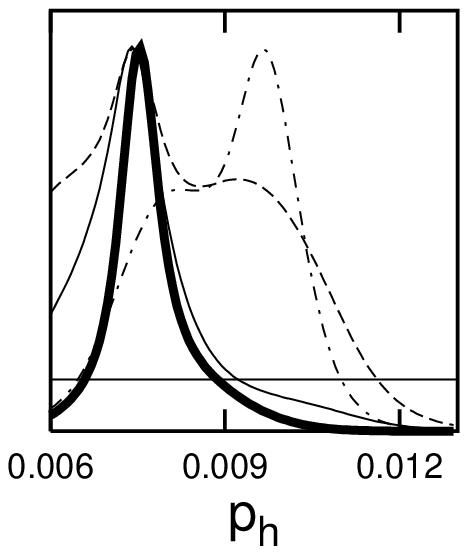}
\epsfxsize=2.7cm
\epsfbox{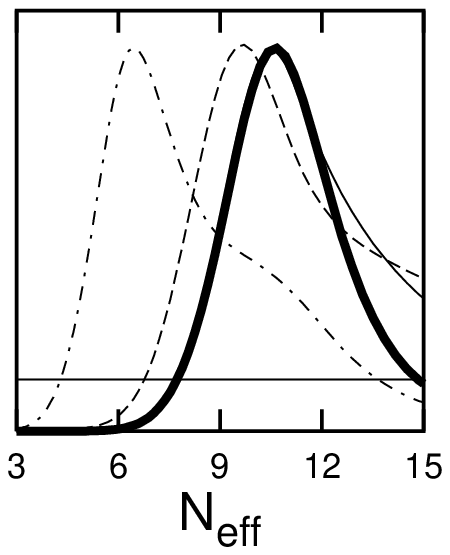}
\epsfxsize=2.7cm
\epsfbox{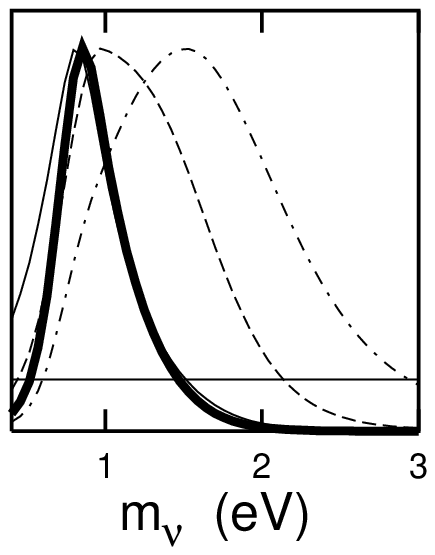}
\\
\caption[fig4]{\label{fig4} The probability distribution for each
cosmological parameter, with one family of massive degenerate
neutrinos.  As in figure~\ref{fig1}, the thick solid line shows the
results from the complete data set, while the thin curves combine the
non-CMB data with each of the three CMB experiments
individually. Maxima and DASI allow a significantly higher baryon
fraction than Boomerang, though the three data sets are compatible.}
\end{figure}

\begin{figure}
\centering 
\leavevmode\epsfysize=4.5cm \epsfbox{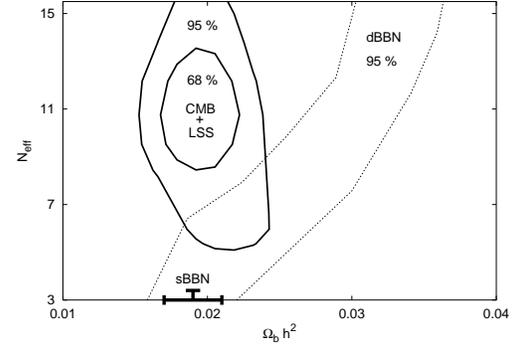}\\ 
\caption[fig5]{\label{fig5} The thick lines show the allowed region in
the ($\Omega_{\rm b} h^2$, $N_{\rm eff}$) parameter space, at 68 and
95 per cent confidence levels, for one family of massive degenerate
neutrinos. On the lower axis we show the standard BBN prediction, and
the thin curves show the region allowed at 95 per cent confidence
level by degenerate BBN (Esposito et al.~2000, 2001). The regions
overlap for $5 \leq N_{{\rm eff}} \leq 9$.}
\end{figure}

The results for the massive case are also given in Table~\ref{tab1},
and the probability for each individual parameter is shown in
figure~\ref{fig4}. The best-fitting model now has an impressively low
$\chi^2$ of 43 and some remarkable features. A large effective
neutrino number between 7 and 15 is preferred, producing a high
first acoustic peak as in $\Lambda$CDM models.  This large lepton
asymmetry is compatible with BBN up to $N_{\rm eff}\simeq9$, as can be
seen in figure~\ref{fig5}. When it is combined with a mass close to
1~eV, it gives the right shape and amplitude for the matter power
spectrum.  A neutrino mass smaller than 0.6~eV is excluded at more
than 95 per cent confidence; this result is in nice agreement with the
oscillations reported at the Los Alamos Liquid Scintillation Neutrino
Detector (Athanassopoulos et al.~1998), which support evidence for a
neutrino mass $m_{\nu}^2 \geq (0.1 - 1) \; {\rm eV}^2$.

Unfortunately, this positive picture is darkened by the predicted
baryon density, which is as low as in the standard case: $\Omega_{\rm
b} h^2 = 0.020^{+0.005}_{-0.004}$ (95 per cent confidence). So, within the
range of viable parameters the lepton asymmetry model can no longer
reach the high baryon fraction potentially allowed by the degenerate
BBN model. Studying the curves for individual CMB experiments in
figure~\ref{fig4}, we see that this result is driven primarily by the
new Boomerang results, with DASI and Maxima both still allowing
significantly higher values; note in particular that inclusion of
neutrino mass allows DASI to go to higher baryon fractions than it can
in the massless case. With all data taken into account, the model is
now restricted to $\Omega_{\rm b} h^{1.5} \leq 0.033$, more than
2-$\sigma$ away from the Ettori \& Fabian (1999) cluster bound.

\section{Summary}

\begin{figure*}
\centering 
\leavevmode
\epsfxsize=7.cm
\epsfbox{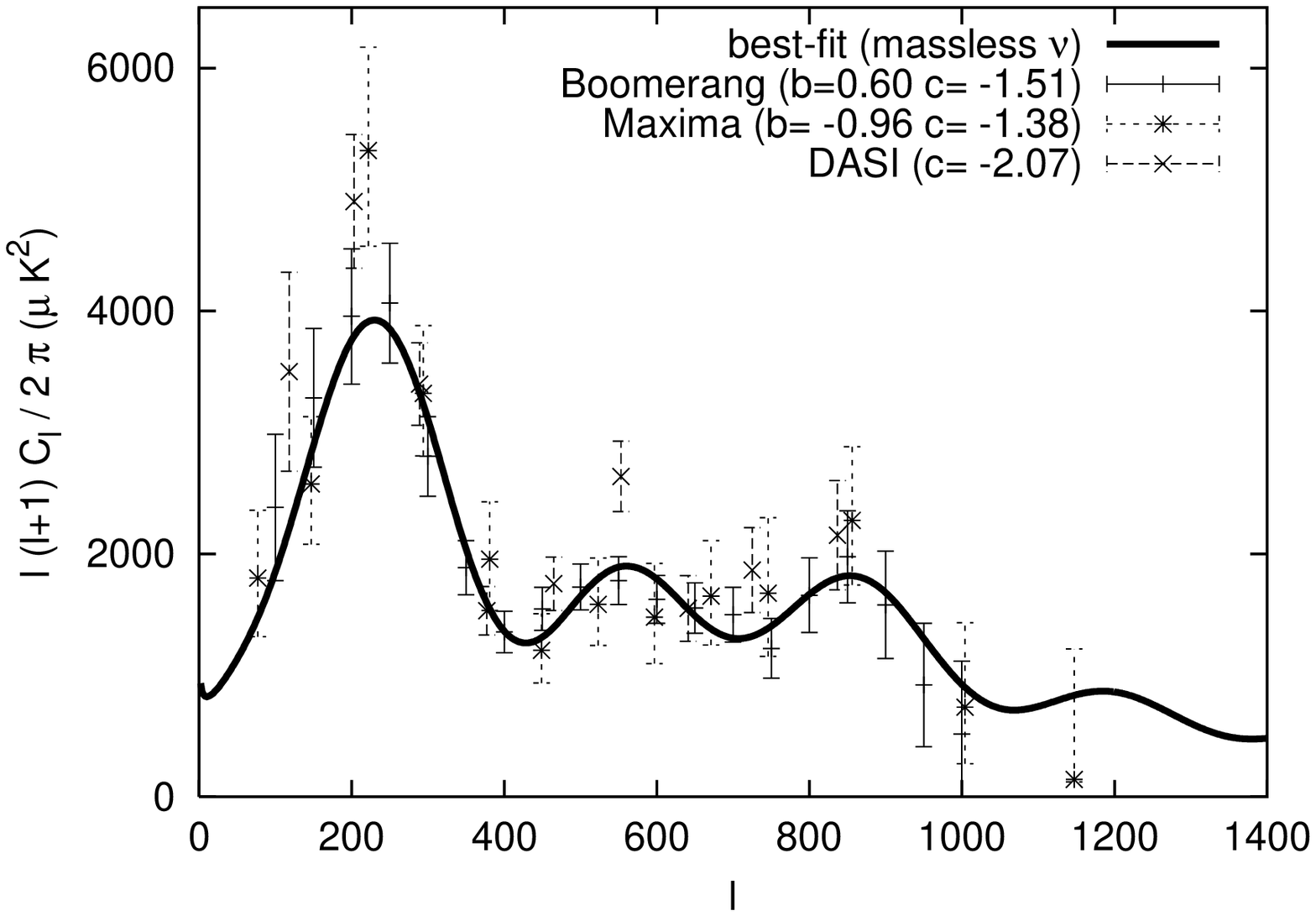}
\epsfxsize=7.cm
\epsfbox{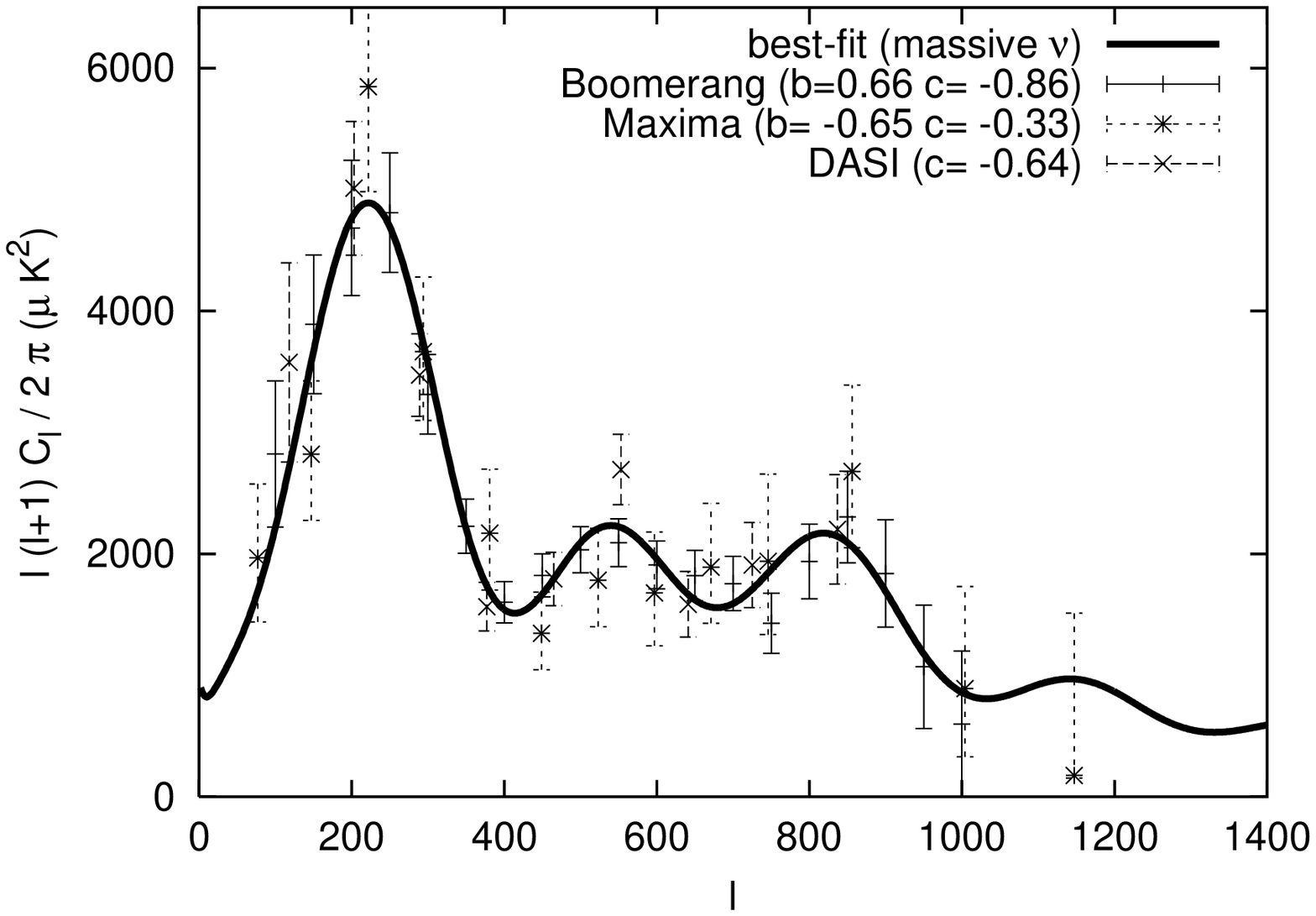}
\vspace{0.5cm}
\\
\leavevmode
\epsfxsize=7.cm
\epsfbox{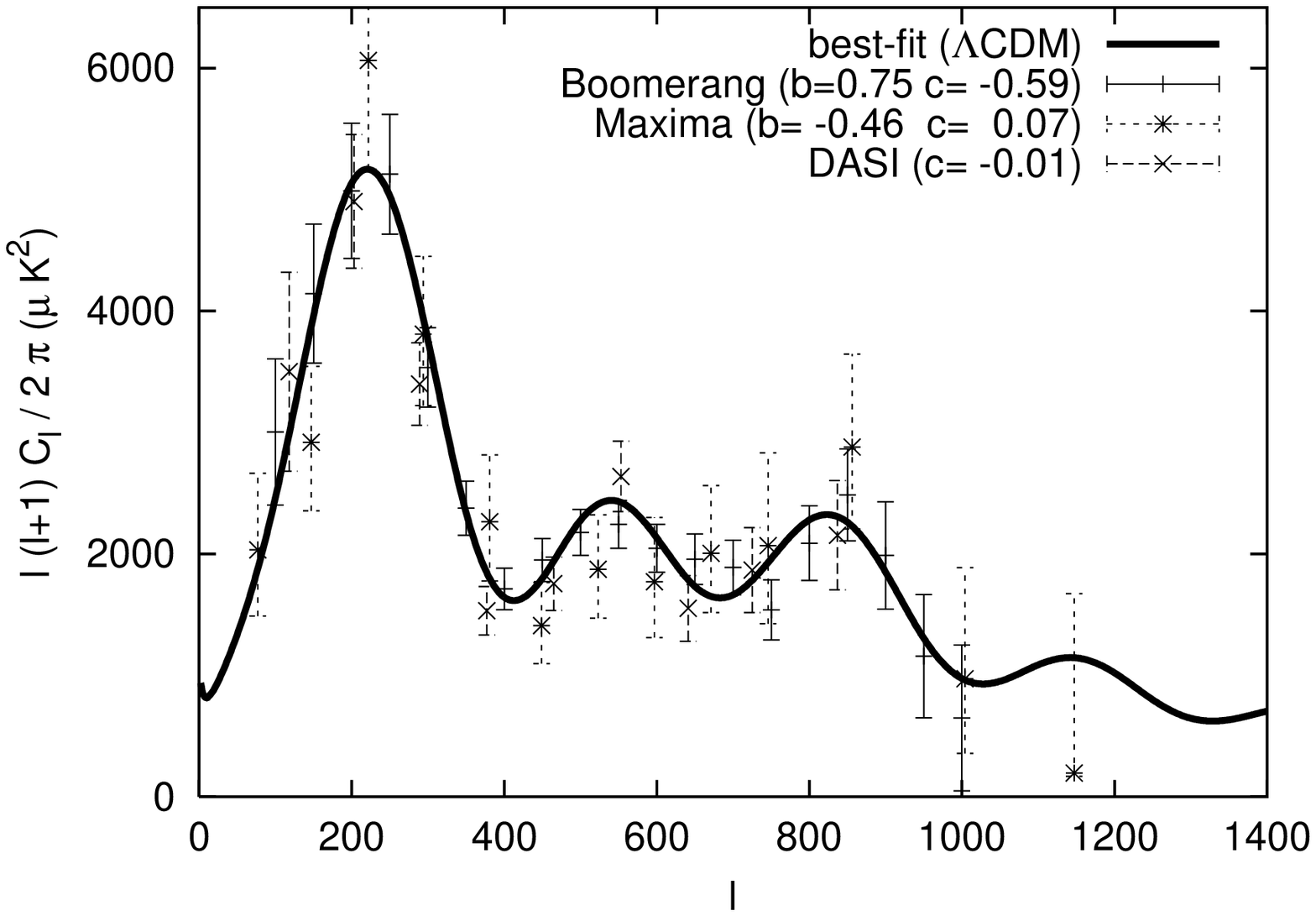}
\epsfxsize=7.cm
\epsfbox{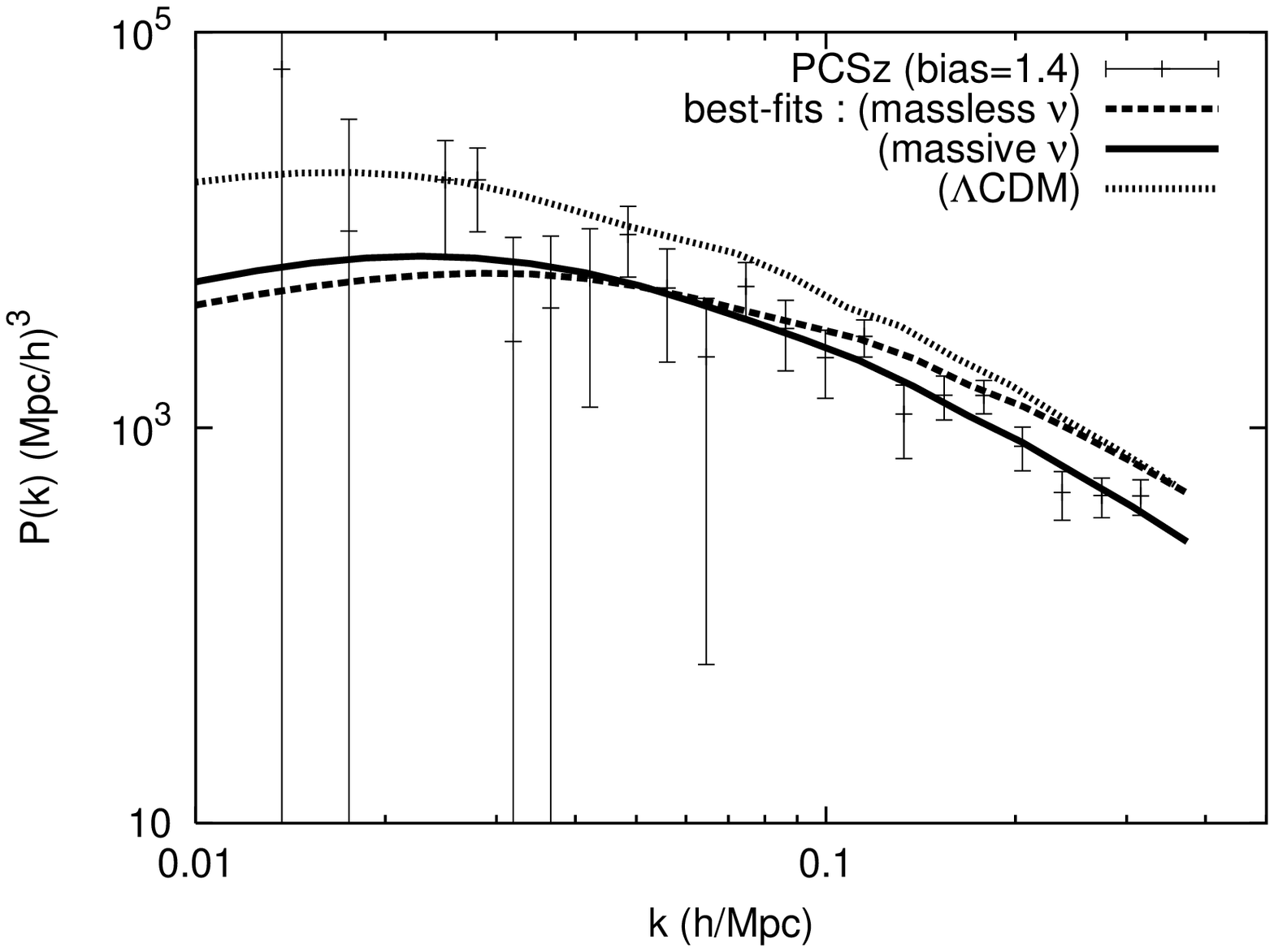}
\caption[fig3]{\label{fig3} The CMB anisotropy and matter power
spectra for three models: the massless and massive degenerate neutrino
best-fitting models (whose parameters are given in Table~\ref{tab1}),
and the preferred $\Lambda$CDM model given by Wang, Tegmark \&
Zaldarriaga~(2001): $\Omega_{\rm b} h^2 = 0.020$, $\Omega_{\rm cdm}
h^2 = 0.012$, $\Omega_{\Lambda}=0.66$, $n_s=0.93$. The corresponding
$\chi^2$ (including CMB and LSS data) are equal to 61, 44 and 56 (the
last is not very good simply because Wang et al. did not include a
$\sigma_8$ constraint in their analysis). In the first three graphs,
the Boomerang, DASI and Maxima data sets are shown with the
appropriate beam and calibration errors $(b, c)$ calculated for each
case. The three matter power spectra are plotted together in the final
plot, along with the PSCz points divided by the square of the bias
factor $b=1.4$ which minimizes the $\chi^2$ for the massive neutrino
model.}
\end{figure*}

We have performed a detailed comparison of critical-density models
including leptonic asymmetry with the latest CMB and LSS data, in
order to investigate their viability as alternatives to the
$\Lambda$CDM model. Some sample power spectra are shown in
figure~\ref{fig3}. We have found that very good fits to those data
are available, due to the combined effect of the large lepton
asymmetry (which is compatible with primordial abundances), and of a
neutrino mass $\sim 1$~eV (which is in nice agreement with LSND). This
model cannot hope to explain the supernovae data, but has the prospect
of undermining the other support for the $\Lambda$CDM paradigm which has
led to its wide acceptance. Unfortunately, the newest CMB data
introduces a new problem for this model, which is that the baryon
density is now constrained to be low enough that fits to the cluster
baryon fraction are not possible, which is disappointing as the lepton
asymmetry model had the potential to allow higher baryon densities
while remaining compatible with nucleosynthesis. The main driving
force to this conclusion is the new analysis of the Boomerang data
(Lee et al.~2001); the other new CMB data still permit a higher baryon
density in the presence of a massive degenerate neutrino. Given the
subtle effects of the neutrino degeneracy, and the various
uncertainties in the new CMB data (calibrations, tilts), this was not
obvious by eye, and it is the main result of this paper.

We stress that our study does not provide a model-independent bound on
the leptonic asymmetry in the Universe, since it could in principle
coexist with a cosmological constant. However, the lepton asymmetry is
better motivated in the critical-density case, with it being used to
remove the need for $\Lambda$.  What our study shows is that following
the recent results, the critical-density lepton asymmetry model
experiences new observational difficulties which make it a less
attractive proposition as a simple and elegant alternative to the
$\Lambda$CDM cosmology.

\section*{ACKNOWLEDGMENTS}

We thank Gianpiero Mangano, Sergio Pastor, Michael Rowan-Robinson and
Subir Sarkar for useful discussions.  J.L.~acknowledges a visit to the
University of Sussex funded in part by PPARC, and ARL the hospitality
of the National Taiwan University where part of this work was carried
out.



\bsp
\end{document}